\documentclass[structabstract]{aa}  
\usepackage{color}
\usepackage{txfonts}
\usepackage[]{graphicx}
\begin{document}
   \title{AKARI's infrared view on nearby stars}

   \subtitle{Using AKARI Infrared Camera All-Sky Survey, 2MASS, and Hipparcos catalogs}

   \author{Yoshifusa Ita\inst{1,2},  Mikako Matsuura\inst{3,4}, Daisuke Ishihara\inst{5}, Shinki Oyabu\inst{2}, Satoshi Takita\inst{2}, Hirokazu Kataza\inst{2}, Issei Yamamura\inst{2}, Noriyuki Matsunaga\inst{6}, Toshihiko Tanab\'{e}\inst{6}, Yoshikazu Nakada\inst{6}, Hideaki Fujiwara\inst{7}, Takehiko Wada\inst{2}, Takashi Onaka\inst{7} \and Hideo Matsuhara\inst{2}}

   \institute{National Astronomical Observatory of Japan, 2-21-1 Osawa, Mitaka, Tokyo, 181-8588 Japan \\
             \email{yoshifusa.ita@nao.ac.jp or yita@ir.isas.jaxa.jp}
         \and
             Institute of Space and Astronautical Science, Japan Aerospace Exploration Agency, 3-1-1 Yoshinodai, Sagamihara, Kanagawa 229-8510, Japan
         \and
             UCL-Institute of Origins, Department of Physics and Astronomy, University College London 
             Gower Street, London WC1E 6BT, United Kingdom
         \and
             UCL-Institute of Origins, Mullard Space Science Laboratory, University College London, Holmbury St. Mary, Dorking, Surrey RH5 6NT, 
             United Kingdom
         \and
             Department of Astrophysics, Nagoya University, Chikusa-ku, Nagoya 464-8602, Japan
         \and
             Institute of Astronomy, School of Science, The University of Tokyo, Mitaka, Tokyo 181-0015, Japan
         \and
             Department of Astronomy, Graduate School of Science, The University of Tokyo, Bunkyo-ku, Tokyo 113-0033, Japan
             }

   \date{Received --, 2009; accepted --, 2009}

   \authorrunning{Ita et al.}

 
  \abstract
   {The AKARI, a Japanese infrared space mission, has performed an All-Sky Survey in six infrared-bands from 9 to 180 $\mu$m with higher spatial resolutions and better sensitivities than the Infrared Astronomical Satellite (IRAS).}
   {We investigate the mid-infrared (9 and 18 $\mu$m) point source catalog (PSC) obtained with the Infrared Camera (IRC) on board the AKARI, in order to understand the infrared nature of the known objects, as well as to identify previously unknown objects.}
   {Color-color diagrams and a color-magnitude diagram have been plotted, using the AKARI-IRC PSC and other available all-sky survey catalogs. We combine the Hipparcos astrometric catalog, and the 2MASS all-sky survey catalog with the AKARI-IRC PSC. We further searched literatures and SIMBAD astronomical database for object types, spectral types and luminosity classes. We identify the locations of representative stars/objects on color-magnitude and color-color diagram scheme. The properties of unclassified sources can be inferred on the basis of their locations on these diagrams.}
   {We found that the ($B-V$) v.s. ($V-S9W$) color-color diagram is useful to identify the stars with infrared excess emerged from circumstellar envelopes/disks. Be stars with infrared excess are well separated from other types of stars in this diagram. Whereas ($J-L18W$) v.s. ($S9W-L18W$) diagram is a powerful tool to classify several object-types.
  Carbon-rich asymptotic giant branch (AGB) stars and OH/IR stars form distinct sequences in this color-color diagram.
  Young stellar objects (YSOs), pre-main sequence (PMS) stars, post-AGB stars and planetary nebulae (PNe) have largest mid-infrared color-excess, and can be identified in infrared catalog.
 Finally, we plot $L18W$ v.s. ($S9W-L18W$) color-magnitude diagram, using the AKARI data together with Hipparcos parallaxes. This diagram can be used to identify low-mass YSOs, as well as AGB stars. We found that this diagram is comparable to the $[24]$ vs $([8.0]-[24])$ diagram of Large Magellanic Cloud sources using the \textit{Spitzer} Space Telescope data. Our understanding of Galactic objects will be used to interpret color-magnitude diagram of stellar populations in nearby galaxies which \textit{Spitzer} Space Telescope has observed. }
   {Our study of the AKARI color-color and color-magnitude will be used to explore properties of unknown objects in future. In addition, our analysis highlights a future key project to understand stellar evolution with circumstellar envelope, once the forthcoming astronometrical data with GAIA are available.}

   \maketitle
%

\section{Introduction}
\begin{figure}
\centering
\includegraphics[angle=0,scale=0.7]{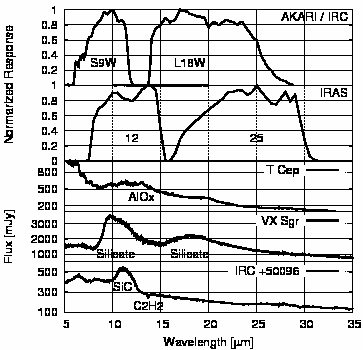}
\caption{The normalized spectral response function of the AKARI IRC bands and the IRAS bands. As references, the ISO SWS spectra of three rep resentative Galactic AGB stars (T Cep as O-rich AGB with AlOx feature, VX Sgr as O-rich AGB with silicate feature, and IRC+50096 as C-rich AGB with SiC feature) with circumstellar dust features are shown.}
\label{filter}
\end{figure}

More than a quarter-century passed since the pioneering infrared whole sky survey of the InfraRed Astronomical Satellite (IRAS), which covered more than 96\% of the whole sky  in four photometric bands at 12, 25, 60 and 100 $\mu$m (IRAS Explanatory Supplement \cite{iras}). The IRAS point source catalog (PSC) has shown that mid- and far-infrared census is essential for studying dust embedded objects, such as star-forming regions, debris disks around main sequence stars, evolved stars, and distant galaxies. However, the spatial resolution was not so good to study sources in the crowded retions. After the IRAS, the COsmic Background Explore (DIRBE/COBE; Hauser et al. \cite{hauser1998}) has mapped the whole sky in 1.25--240\,$\mu$m with 10 photometric-bands. It intended to accurately obtain the intensity of diffuse background radiation and did not have high sensitivity for point sources. In the meanwhile, ultra-violet, optical and near-infrared large area surveys have carried out using ground based telescopes (e.g. GALEX (Martin et al. \cite{martin2005}); SDSS (York et al. \cite{york2000}); 2MASS (Skrutskie et al. \cite{skrutskie2006}). Their counterparts are missing or hard to be uniquely identified in mid-infrared and far-infrared catalogs, which prevents us from studying objects surrounded by dust. The demand of new mid- and far-infrared whole sky survey with better sensitivity and higher spatial resolution has been increased. To fulfill the expectations, the AKARI, a Japanese infrared satellite was launched at 21:28 UTC on February 21st, 2006 from the Uchinoura Space Center (Murakami et al. \cite{murakami2007}). Sharing the time with the pointed observations, AKARI has mapped the whole sky in mid- and far-infrared using two instruments on board; the InfraRed Camera (IRC; Onaka et al. \cite{onaka2007}) and the Far-Infrared Surveyor (FIS; Kawada et al.  \cite{kawada2007}). The FIS swept about 94\% of the whole sky more than twice at  65, 90, 140, and 160 $\mu$m wavebands.  Also, the IRC swept more than 90\% of the whole sky more than twice using two filter bands centered at 9 ($S9W$, 7 -- 12 $\mu$m) and 18 ($L18W$, 14 -- 25 $\mu$m) $\mu$m (Ishihara et al. \cite{ishihara2010}). These abbreviated filter band names are used throughout this article. The cut-in and cut-off wavelengths indicated in the parentheses correspond to those where the throughput becomes a half of the peak. See Figure~\ref{filter} for the normalized spectral response function of the IRC bands.

In this paper, we use the IRC mid-infrared All-Sky Survey data to study Galactic stellar objects. Compared to the IRAS survey, the sensitivities at 9 and 18 $\mu$m bands are more than 15 and 5 times better than those of the IRAS's 12 and 24 $\mu$m bands, and the spatial resolution is more than 100 times finer for the IRC survey. van der Veen \& Habing (\cite{veen1988}) utilized IRAS mid- and far-infrared combined colors as a tool to diagnose the nature of IRAS sources. They defined color criteria to classify sources into several groups, and since then those criteria have been used by many authors. Taking this as a role model, we cross-identify the AKARI IRC All-Sky Survey point source catalog with the Hipparcos astrometric catalog (van Leeuwen \cite{vanLeeuwen2007}), and the 2MASS all-sky survey catalog (Skrutskie et al. \cite{skrutskie2006}), to classify sources on color-color and color-magnitude diagrams. The main objective of this paper is to study overall infrared characteristics of galactic stellar sources. The new All-Sky Survey catalog should be useful for wide variety of astronomical studies. Practical applications of the catalog are, search for hot debris disk (Fujiwara et al. \cite{fujiwara2010}), extragalactic objects (Oyabu et al. in preparation) and study on the young stellar objects (Takita et al. \cite{takita2010}). Refer to the papers for discussions on each topic.

In the next section we show general characteristics of the AKARI IRC All-Sky Survey point source catalog. Refer to Ishihara et al. (\cite{ishihara2010}) for the complete description of the All-Sky Survey, its data reduction processes, the point source catalog compilation processes, and the catalog characteristics.



\begin{figure}
\centering
\includegraphics[angle=-90,scale=0.36]{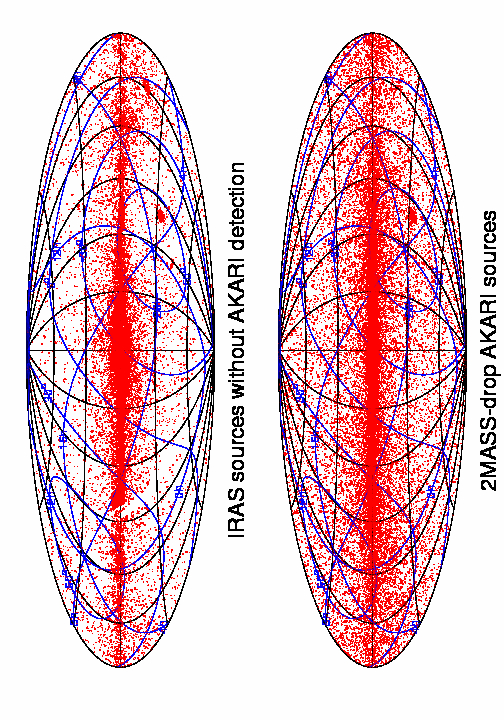}
\caption{A projection of the IRAS sources without AKARI detection (upper panel) and the 2MASS-drop AKARI sources (lower panel) onto the galactic coordinate map. The blue lines show the equatorial coordinate.}
\label{galdist}
\end{figure}

\section{Characteristics of the AKARI IRC All-Sky Survey catalog}
The first release version (ver. $\beta$-1) of the AKARI IRC mid-infrared All-Sky Survey point source catalog (hereafter, we call the catalog as IRC-PSC) lists more than 851,000 and 195,000 sources at 9 and 18 $\mu$m bands, respectively. There are about 170,000 sources detected both in the two bands. This number is in a range expected from the difference in the dection limit between the two bands and the fall of the Rayleigh-Jeans spectrum since most of the sources are stars. The estimated 5 $\sigma$ detection limits for one scan are about 50 and 90 mJy at 9 and 18 $\mu$m bands, respectively. The present catalog includes point-like sources that were detected more than twice. The sensitivity will be improved in the future catalog, for sources in high visibility regions for AKARI's sun-synchronous orbit (i.e., high ecliptic latitude regions), where AKARI scanned many times. The saturation limits depend on the sky region and the brightest source listed in the catalog is about 560 and 1,200 Jy at 9 and 18 $\mu$m bands, respectively. The pixel field of view of the survey observation mode is about 10 arcsec, and the positional accuracy of detected sources are better than 3 arcsec.

\subsection{Flux accuracy}
\subsubsection{Comparison to the IRAS point source catalog}
Although the band profiles of AKARI IRC's $S9W$ and $L18W$ bands and IRAS's 12 and 25 $\mu$m bands are different, a comparison of the photometry of common sources is useful to test the calibration of the IRC-PSC. In Figure~\ref{filter} we show the normalized spectral response function of the AKARI and the IRAS\footnote{data taken from \\ http://irsa.ipac.caltech.edu/IRASdocs/exp.sup/ch2/tabC5.html} bands. The ISO SWS spectra (Sloan et al. \cite{sloan2003a}) of some galactic stars with characteristic circumstellar dust features are also shown to get the rough idea on the cause of differences in photometry between the associated filter bands.

The IRAS-PSC lists 245,888 sources, among which 170,754 have flux quality flag better than 1 in at least one of the 12 and 25 $\mu$m bands (i.e., $f_{q12} > 1$ or $f_{q25} > 1$). We refer these 170,754 sources as good IRAS sources. We find AKARI counterparts for 145,751 ($> 85 \%$) good IRAS sources using the simple positional matching method with a tolerance radius of 30 arcsec. In some cases, more than one AKARI point sources are found for an IRAS source. In these cases, we only adopt the closest one and regard the other(s) as unmatched, even if they are actual multiple sources resolved by the AKARI that appear as one source to the IRAS.  We compare the AKARI and IRAS photometry of the matched sources. The comparison shows that the photometry in the IRC-PSC and the IRAS-PSC are in agreement within 37 and 40\% in S9W v.s. IRAS12, and L18W v.s. IRAS25 for sources with the IRAS flux quality flag of 3. If we compare a subsample of high galactic latitude ($|b| > 30^\circ$) and high quality (S/N $>$ 10 in IRC bands) sources, their photometry are in agreement within 18 and 24\% in S9W v.s. IRAS12, and L18W v.s. IRAS25, respectively (see also Ishihara et al., \cite{ishihara2010}). The comparison also reveals that there are about 25,000 good IRAS sources that are without AKARI counterparts. Their galactic spatial distribution is shown in the upper panel of the Figure~\ref{galdist}. There are several possibilities to explain why some of the IRAS sources (especially bright ones) are not listed in the IRC-PSC. The possible explanations are, (1) saturated in the AKARI survey  (2) the IRAS point-like source can be recognized as an extended source to the AKARI's eye, and excluded from the "point source" catalog, (3) the IRAS source can be resolved into several fainter sources to the AKARI's eye,  (4) difference in the sky coverage ($\sim$ 90\% for AKARI and $\sim$ 96\% for IRAS) or located in the unexplored sky region (Ishihara et al. \cite{ishihara2010}).

\subsection{Positional accuracy}
The 2MASS point source catalog (2MASS-PSC; Skrutskie et al. \cite{skrutskie2006}) is complete down to $K_s < 14.3$ mag in the absence of confusion\footnote{http://www.ipac.caltech.edu/2mass/releases/allsky/doc/sec2\_2.html}. Then, all normal stars seen by AKARI should be prominent in the 2MASS-PSC. Therefore, we use the 2MASS-PSC to assess the positional accuracy of the IRC-PSC.

There are 761,565 sources with S/N $>$ 5 at the $S9W$ band in the IRC-PSC. We refer these 761,565 sources as good IRC sources. We search the 2MASS counterparts brighter than 14.3 mag and S/N $>$ 5 in $K_s$ band for the good IRC sources. If more than one 2MASS sources are found within the tolerance radius from an AKARI source, we only adopt the closest one and regard the other(s) as unmatched. We find counterpart for 505,350 (66\%), 673,730 (88\%), 713,705 (94\%), 724,739 (95\%), and 728,178 (96\%) good IRC sources using the positional tolerance radii of 1, 2, 3, 4, and 5 arcsec, respectively. If we use a subsample of 562,598 high quality (S/N $> 10$ at the $S9W$ band) IRC-PSC sources, the results would be 68\%, 88\%, 93\%, 94\%, and 94\% for the positional tolerance radii of 1, 2, 3, 4, and 5 arcsec, respectively. These results suggest that the positional accuracy of the IRC-PSC does not depend on the source brightness, and the accuracy is uniform at least for sources with S/N $>$ 5. Also, it seems that the chance of false matches may get large if we use the tolerance radius larger than 4 arcsec.

To see the dependence of positional accuracy on the source density, we make the same 2MASS counterpart search with a subsample of 78,171 high galactic latitude ($|b| > 30^\circ$) good IRC-PSC sources, where we do not suffer from severe confusion. We search the 2MASS counterpart brighter than 14.3 mag and S/N $>$ 5 in $K_s$ band for these high latitude good IRC-PSC sources, and find counterpart for 61,619 (79\%), 70,446 (90\%), 71,877 (92\%), 72,334 (93\%), and 72,528 (93\%) sources using the positional tolerance radii of 1, 2, 3, 4, and 5 arcsec, respectively. This result indicates that the positional accuracy of the IRC-PSC may depend on the source density, but it does not matter if we use the match radius larger than 2 arcsec.

Considering the above test results, we conclude that the positional accuracy of the IRC-PSC sources is better than 3 arcsec for most of the sources. This result is compatible with the pointing accuracy estimated in Ishihara et al. (\cite{ishihara2010}). Then we decide to use a tolerance radius of 3 arcsec for catalog comparisons made in the next section.

Finally, the AKARI sources without 2MASS counterpart should be of particular interest because such sources can be deeply dust enshrouded objects (e.g., OH/IR stars, dusty carbon stars, etc.) or distant galaxies. There are 47,860 good IRC sources without 2MASS counterpart (brighter than 14.3 mag and S/N $>$ 5 in $K_s$ band) within a radius of 3 arcsec. We show the galactic spatial distribution of these 2MASS-drop AKARI sources in the lower panel of the Figure~\ref{galdist}. Follow-up observations are definitely needed to identify these sources.

\begin{figure}
\centering
\includegraphics[angle=-90,scale=0.36]{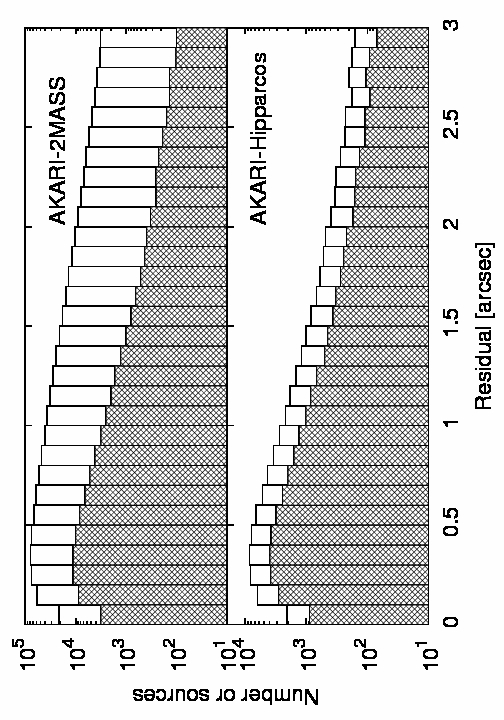}
\caption{The histogram of positional differences between the AKARI IRC PSC and the 2MASS catalog (upper panel), and the Hipparcos catalog (lower panel) for matched sources. The hatched areas represent matched sources in the high galactic latitude ($|b| > 30^\circ$) region. The size of the bin is 0.1 arcsec.}
\label{histogram}
\end{figure}


\section{Cross-identification with existing catalogs/databases and data set definition}
The IRC-PSC is cross-identified with existing all-sky survey catalogs, namely the new Hipparcos astrometric catalog (van Leeuwen \cite{vanLeeuwen2007}) and the 2MASS PSC (Skrutskie et al. \cite{skrutskie2006}), using a simple positional correlation method. The epoch of the source positions listed in the new Hipparcos catalog is 1991.25, while that of the IRC-PSC is 2000.0. There are 15,052 Hipparcos sources whose total proper motions are larger than 100 milliarcsec/year. Their positions should be corrected for the proper motion over the 8.75 year interval. We calculate the positions in epoch 2000.0 for the Hipparcos sources with good proper motion measurements ($\sigma \mu_\alpha/\mu_\alpha < 0.2$ and $\sigma \mu_\delta/\mu_\delta < 0.2$, where $\mu_\alpha$ and $\mu_\delta$ are proper motions in right ascension and declination, respectively). Then we use the new positions for the cross-identification.

We use a positional tolerance of 3 arcsec amid the positional accuracy of the IRC-PSC ($<$ 3 arcsec). If more than one sources are present within the tolerance radius, the closest one is selected. In Figure~\ref{histogram}, we show the histogram of positional differences for matched sources. We find 68,744 matches between the IRC-PSC and new Hipparcos catalog, and 847,838 matches between the IRC-PSC and the 2MASS catalog within the tolerance radius of 3 arcsec. The optical and near-infrared photometries used in this article are taken from these catalogs. Magnitudes are not corrected for interstellar extinction. Instead, we will provide an indication of interstellar extinction by showing extinction vector. We use Weingartner \& Draine (\cite{weingartner2001}) extinction model for the Milky Way of $R_v = 3.1$ to calculate extinction vectors. We take improved astrometry, ($B-V$) colors and their errors, and ($V-I$) colors from van Leeuwen (\cite{vanLeeuwen2007}), and $V$-band magnitudes are extracted from older version of the Hipparcos catalogue (ESA, \cite{esa1997}). We assume that the errors in $V$ magnitudes are $\sigma (V) = \sigma(B-V)/\sqrt2$.

\subsection{Sources with known classifications}
After the cross-identification processes, we search astronomical catalogs that compile interesting types of objects. Then we cross-identify our data with the following astronomical catalogs using a tolerance radius of 3 arcsec; (1) S-type stars (stars with surface carbon-to-oxygen number ratio (C/O ratio) close to the unity):  Stephenson (\cite{stephenson1984}, \cite{stephenson1990}), who list 1,412 sources, (2) Post-AGB stars: Szczerba et al. (\cite{szczerba2007}), who list 326 very likely post-AGB stars, (3) Planetary Nebulae: Acker et al. (\cite{acker1994}), who list 1,143 sources, (4) Wolf-Rayet stars: van der Hucht (\cite{hucht2001}), who list 226 sources, (5) Be stars:  Zhang et al (\cite{zhang2005}), who list 1,185 sources. We further search the SIMBAD database for known carbon stars, OH/IR stars, pre-main-sequence (PMS) stars, and young stellar objects (YSOs), by querying by object-types of 'C*' , 'OH*', 'pMS*', and 'Y*O', respectively. Also, M-type giants and supergiants are extracted from the SIMBAD database by querying by spectral types of 'M' and luminosity class of 'III (giants)' and 'I (supergiants)', respectively. We refer the classifications in the above mentioned literatures and the SIMBAD database query results throughout this article unless otherwise noticed. Occasionally, more than one object-types are found (within a search radius of 3 arcsec) for a single object in the SIMBAD database. In those cases, we keep both of types. 

As for the SIMBAD query results, we eliminate nine sources from the PMS stars list and one source (HD217086) from the YSOs list. The nine PMS stars are eliminated because these stars are classified as PMS stars by citing Li \& Hu (\cite{li1998}), but there is only one star (HD21051) among the nine sources that Li \& Hu (\cite{li1998}) indeed identified as a "candidate" weak-lined T Tauri star. Other eight sources are just listed in their paper as possible optical counterparts to the ROSAT All-Sky Survey Bright Source Catalog (Voges et al. \cite{voges1996}). Also, a literature search made us colclude that HD217086 is a O7 dwarf, not a YSO. These ten sources are tabulated in Table~\ref{eliminated} with their names, adopted object-types, and references. After this manual procedure, the number of sources in our data set with known object-types/spectral types are summarized in Table~\ref{objtype}. The electronic tables for these data sets are available at http://somewhere. Tables~\ref{akari-hipparcos} and \ref{akari-2mass} are the examples of the data sets provided, and show the first 3 records of them.

\subsubsection{Extragalactic objects}
Extragalactic objects can be contaminants when studying galactic objects in the infrared. As of January 2010, the NASA Extragalactic Database (NED) lists 2,907 classified extragalactic objects with the IRAS 12 $\mu$m flux brighter than 100 mJy. We match these NED objects with the IRC-PSC and find 794 matches within a tolerance radius of 3 arcsec. This matching result indicates that most of the bright extragalactic objects are identified as extended source by AKARI, and only distant ones that appear point-like are included in the IRC-PSC. These 794 objects are eliminated from the following analyses. We further investigate contaminant of extragalactic objects in high Galactic latiude regions.  There are 85,965 AKARI sources detected at either of the S9W or L18W band in the region of $| b | > 30^\circ$. We search the NED database for galaxies or QSOs within a radius of 3 arcsec from those AKARI sources and obtain 1224 matches (1.4\%). Among the 1,224 sources, 412 have the IRAS 12 $\mu$m flux brighter than 100mJy. This is a sufficiently small number and we conclude that the extragalactic contaminant has little or insignificant effects on the following analyses.


\begin{table}
\caption{Sources eliminated from the SIMBAD's pre-main-sequence list.}  
\label{eliminated}
\begin{center}
\begin{tabular}{l l r}
\hline\hline
Name & Adopted object-type & Reference \\ 
\hline
VY Ari       & Late type stars in binary system & 1 \\   
V573 Per  & G8 sub-giant  & 2 \\
HD 21051 & K0 giant  & 2  \\
UX Ari       & Late type stars in binary system & 3 \\
46 Per      & Emission line OB stars & 4 \\
V491 Per  & G8 sub-giant & 2 \\
V582 Per  & F7 dwarf & 5 \\
V492 Per  & K1 giant in binary system & 6 \\
111 Tau    & F8 dwarf & 7 \\
HD 217086 & O7 dwarf & 8 \\
\hline
\end{tabular}
\end{center}
References:
1: Biazzo et al. (\cite{biazzo2006}), 2: De Mediros \& Mayor (\cite{medeiros1999}), 3: Rosario et al. (\cite{rosario2007}), 4: Puls et al., (\cite{puls2006}), 5: Busa et al. (\cite{busa2007}), 6: De Medeiros et al. (\cite{medeiros2002}), 7: Cenarro et al., (\cite{cenarro2007}), 8; Mokiem et al., (\cite{mokiem2005})
\end{table}

\begin{table}
\caption{Number of sources in our data set with known object-types or spectral-types.}  
\label{objtype}
\begin{center}
\begin{tabular}{l r r}
\hline\hline
Object types /  & \multicolumn{2}{c}{Data set} \\ 
Spectral types & AKARI-Hipparcos & AKARI-2MASS \\ 
\hline
Be stars                              &   449  &   641 \\
Carbon stars                      &   281  & 4606 \\
M-type supergiants            &     74  &   229 \\
M-type giants                     & 1732  & 3129 \\
OH/IR stars                        &       1  &   570 \\
Planetary Nebulae             &      7   &   275 \\
Post-AGB  stars                 &     42  &   242 \\
S-type stars                       &     74  & 1229 \\
Wolf-Rayet stars                &     54  &   100 \\
Young stellar objects         &      8   &  788 \\
pre-main-sequence stars  &     65  &   351 \\
\hline
\end{tabular}
\end{center}
\end{table}

\begin{table*}
\caption{The first three records of AKARI-Hipparcos samples with known object types. The sources are sorted by increasing order of R.A. for each object type. Any indices with prefix "e" mean their errors. The full version of the table is available at http://somewhere.}  
\label{akari-hipparcos}
\begin{center}
\begin{tabular}{r r r r r r r r r r r r r}
\hline\hline
\multicolumn{2}{c}{AKARI coordinate} & $S9W$ & e$S9W$ & $L18W$ & e$L18W$ & Hipparcos & plx & eplx & $V$ & $(B-V)$ & e$(B-V)$ & Object \\
\multicolumn{2}{c}{R.A. \& DEC [degree]} & \multicolumn{4}{c}{[vega-mag]} & ID & \multicolumn{2}{c}{[mas]} & \multicolumn{3}{c}{[vega-mag]} & Type \\
\hline
   0.86307 &  55.55097  & 5.733  & 0.037 & 99.999 & 99.999 &   278  &  1.57   & 0.82  & 7.940 & -0.023 & 0.003  &   Be \\
   1.61056 &  64.19614  & 5.424  & 0.026 & 99.999 & 99.999 &   531  &  3.76   & 0.21  & 5.570 & -0.023 & 0.003  &   Be \\
   2.90472 &  58.21186  & 5.457  & 0.024 &   4.592 &   0.103 &   940  &  0.94   & 0.51  & 7.090 &  0.137 & 0.009  &   Be \\
\hline
\end{tabular}
\end{center}
\end{table*}

\begin{table*}
\caption{The first three records of AKARI-2MASS samples with known object types. The sources are sorted by increasing order of R.A. for each object type. Any indices with prefix "e" mean their errors. The full version of the table is available at http://somewhere.}  
\label{akari-2mass}
\begin{center}
\begin{tabular}{r r r r r r r r r r r r r}
\hline\hline
\multicolumn{2}{c}{AKARI coordinate} & $S9W$ & e$S9W$ & $L18W$ & e$L18W$ & $J$ & e$J$ & $H$ & e$H$ & $K_s$ & e$K_s$ & Object \\
\multicolumn{2}{c}{R.A. \& DEC [degree]} & \multicolumn{10}{c}{[vega-mag]} & Type \\
\hline
   0.35326 &  63.50464  & 7.367  & 0.157 & 99.999 & 99.999 & 99.999 & 99.999 & 99.999 & 99.999  & 7.978  & 0.016 &    Be \\
   0.86307 &  55.55097  & 5.733  & 0.037 & 99.999 & 99.999 &   7.266 &   0.018 &   7.225 &   0.046  & 7.090  & 0.020 &    Be \\
   1.61056 &  64.19614  & 5.424  & 0.026 & 99.999 & 99.999 &   5.463 &   0.037 &   5.470 &   0.044  & 5.468  & 0.023 &    Be \\
\hline
\end{tabular}
\end{center}
\end{table*}

\begin{figure*}
\centering
\includegraphics[angle=-90,scale=0.74]{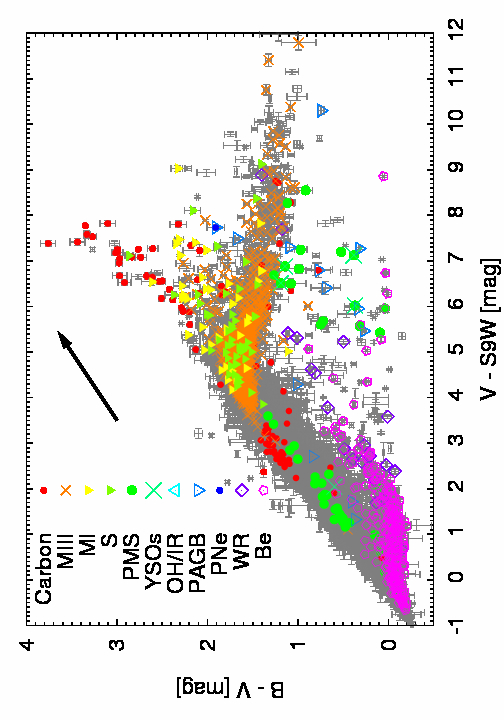}
\caption{The $(V-S9W)$ v.s. $(B-V)$ color-color diagram. The error bars show $\pm$1 $\sigma$ errors in colors. Only sources with S/N $>$ 5 in both colors are plotted. There are 58,793 sources that match the criteria. Symbols indicate object-types as described in the figure, and grey dots (with error bars) show colors of objects without known identifications. Carbon, MIII, MI, S, PMS, YSOs, OH/IR, PAGB, PNe, WR, and Be represent carbon stars, M-type giants, M-type supergiants, S-type stars, pre-main sequence stars, young stellar objects, OH/IR stars, post-AGB stars, planetary nebulae, Wolf-Rayet stars, and Be stars, respectively. Occasionally, more than one object-types are found for a single object in the SIMBAD database. In that case, we show all of the object-types. The black arrow shows interstellar extinction vector for $A_v = 2$ mag, using Weingartner \& Draine (\cite{weingartner2001}) Milky Way model of $R_v = 3.1$.}
\label{VS9W-BV}
\end{figure*}

\section{Results and Analyses}
\subsection{Color-color diagrams}
\subsubsection{ ($V-S9W$) v.s. ($B-V$)}
Figure~\ref{VS9W-BV} shows the ($V-S9W$) v.s. ($B-V$) color-color diagram for AKARI-Hipparcos nearby sources. Only sources with S/N $>$ 5 in both colors are plotted in the figure. There are 58,793 sources that match the criteria.

\paragraph{\bf{Be stars and Wolf-Rayet stars}\\}
In the color-color diagram, there is a distinct sequence consisting of Wolf-Rayet stars and Be stars, starting at about $(B-V) \sim -0.2$ and $(V-S9W) \sim 0$ and extending towards $(B-V) \sim 0.5$ and $(V-S9W) \sim 3$, up to $(V-S9W) \sim 9$. Some of these types of stars clearly show strong infrared excess in $S9W$-band than ordinary B- or O-type stars. Thermal emission from dust grains contributes to this infrared excess (Waters et al. \cite{waters1987}; Zhang et al., \cite{zhang2004}) in addition to free-free and bound-free radiation from the surrounding nebulosity (e.g.,  Wright \& Barlow  \cite{wright1975}; Chokshi \& Cohen \cite{chokshi1987}). In our plot, objects with circumstellar envelope include the Herbig Ae/Be star HD100546 and the Be star, $\iota$\,Ara, which have been identifed as Be stars with infrared excess (Waters et al. \cite{waters1987}). There are several objects with similar colors without identifications, suggesting potential candidates of Wolf-Rayet stars and Be stars with infrared excesses.

\begin{figure}
\centering
\includegraphics[angle=-90,scale=0.36]{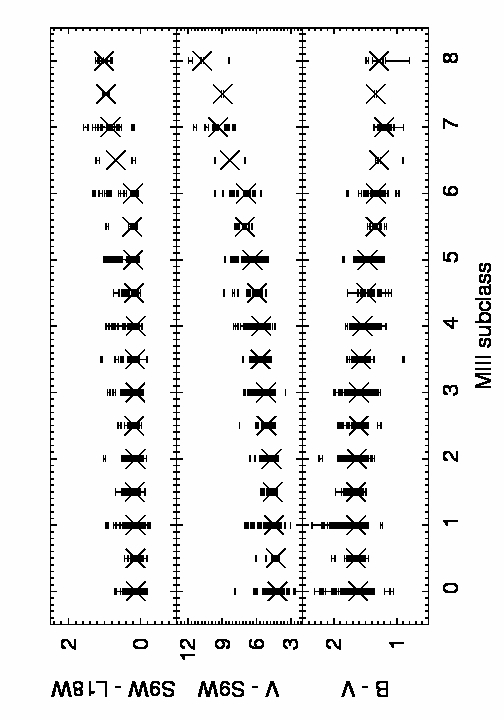}
\caption{Correlation between colors and spectral indexes of M-type giants. Only sources with S/N $>$ 5 in each color are shown. The error bars show $\pm$1 $\sigma$ errors in colors. The big crosses show median values of the colors of each spectral index.}
\label{spvscolor}
\end{figure}

\paragraph{\bf{M-type giants}\\}
M-type giants and carbon stars are clearly separated in the diagram. M-type giants have a decreasing $(B-V)$, as  $(V-S9W)$ becomes redder, while the majority of carbon stars have an increasing trend in both $(B-V)$ and $(V-S9W)$, although there is some scatter amongst carbon stars. This color-color diagram is useful to separate two chemical types of asymptotic giant branch (AGB) stars.

M-type giants have decreasing $(B-V)$, as  $(V-S9W)$ becomes redder, because of the anti-correlation of $(B-V)$ and spectral indices of M-type giants. Figure~\ref{spvscolor} shows the relations between the colors and the spectral subclass indexes of M-type giants. The spectral indices may have a range, such as M3/M4III due to time variations of spectral types, and we took the mean values in such cases. It is clear that $(B-V)$ and $(V-S9W)$ colors show a different response to the increase in spectral index, such that $(B-V)$ becomes bluer with increasing M spectral index, while $(V-S9W)$ becomes redder gradually. By definition, the spectral subclass indices represent the effective temperature ($T_{\textrm{eff}}$) of the star. Based on the model atmosphere, Bessel et al. (\cite{bessell1998}) showed that the $(B-V)$ color of M-type giants anti-correlates with $T_{\textrm{eff}}$. This trend is found in stars with solar metallicity or above, while a linear-correlation is found for sub-solar metallicity stars. Andrews (\cite{andrews1975}) suggested that, for M-type giants, the suppression of fluxes due to TiO molecules is not as strong in $B$-band as in the $V$-band. As $T_{\textrm{eff}}$ decreases, the TiO band absorption is expected to become stronger, and suppress the $V$-band fluxe of M-type giants. It thus seems that the TiO absorption can explain the trend of M-type giants being bluer in $(B-V)$ colors with increasing spectral indices.

\paragraph{\bf{M-type supergiants}\\}
There are six M-type supergiants isolated around $7 < (V-S9W) < 9.5$ and $2 < (B-V)$, and additional three stars are found in the similar color-region, which can be M-type supergiants. These M-type supergiants have much redder $(B-V)$ colors than those of M-type giants at a given $(V-S9W)$ color. What makes M-type supergiants redder in $(B-V)$ color than those of M-type giants? Interstellar extinction may play some role. However, other early-type stars (i.e., O, and B stars, which locate around ($V-S9W$, $B-V$) $\sim$ (0, 0)), who can be also affected by their nearby clouds, seem not to be reddened so much. There should be other reasons. One possible explanation is that the effective surface temperatures of M-type supergiants are generally lower than those of M-type giants, if compared with the same spectral type. For example, Cox (\cite{cox2000}) lists $T_{\textrm{eff}}$ of 3380 K for the M5 giant, but 2880 K for the M5 supergiant. However, lower $T_{\textrm{eff}}$ tends to make $(B-V)$ color of M-type giants bluer as discussed above, and one could imagine that the situation is almost the same for M-type supergiants. This contradiction may be understood by the difference of surface gas density between giants and supergiants. The surface gas density of M-type supergiants is generally lower than that of M-type giants (Cox \cite{cox2000}). Therefore, at a given $T_{\textrm{eff}}$, M-type supergiants should be less effective in forming molecules (e.g., TiO) that contribute to absorb $V$-band flux. The other possibility is due to line blanketing in the $B$-band. Massey (\cite{massey2002}) wrote, \textit{"At a given ($V-R$) color, a low-gravity star will have a larger $(B-V)$ value than a higher gravity star, as a result of the increased importance of line blanketing at lower surface gravities, which is most pronounced in the $B$-band because of the multitude of weak metal lines in the region"}, and as one might expect, the surface gravity of the M-type supergiant is lower by more than 1 dex than that of the M-type giant (e.g., Cox \cite{cox2000}). The influence of the circumstellar dust is also a possibility. The separation of M-type supergiants from M-type giants can be a result from the combination of these factors. Some M-type supergiants seem to overlap with M-type giants in the Figure~\ref{VS9W-BV}. We suspect that they are miss-identified, and are acutually M-type giants, not supergiants (See section 3.2 also).


\begin{figure*}
\centering
\includegraphics[angle=-90,scale=0.68]{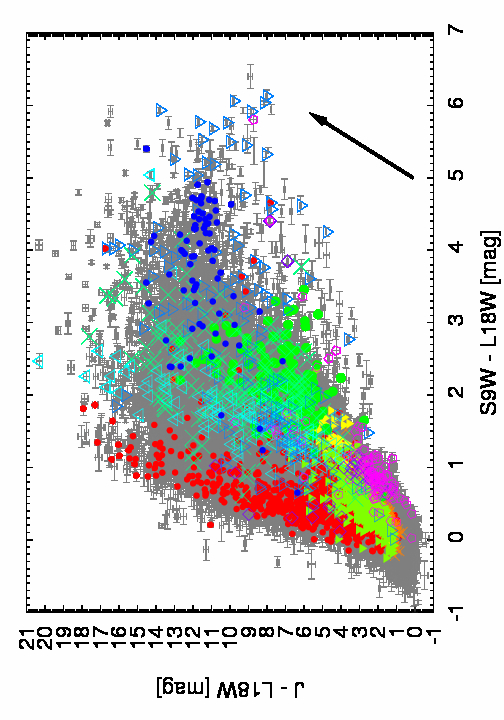}
\includegraphics[angle=-90,scale=0.65]{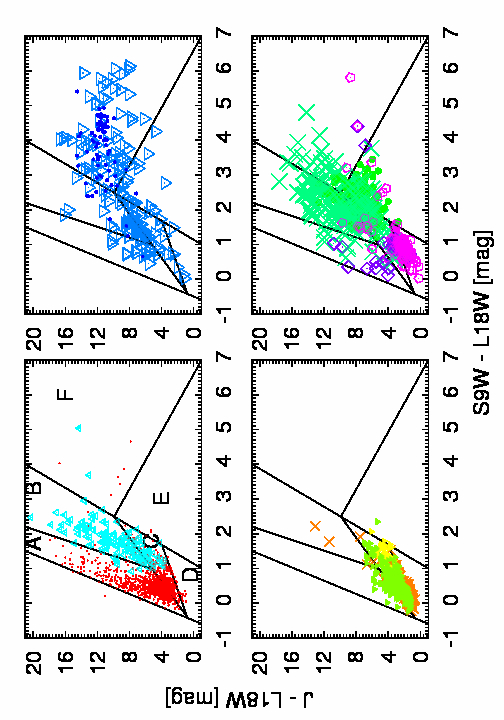}
\caption{Upper panel: The $(S9W-L18W)$ v.s. ($J-L18W$) color-color diagram. The error bars show $\pm$1 $\sigma$ errors in colors. Only sources with S/N $>$ 5 in both colors are included. There are 117,576 sources that match the criteria. Symbols are the same as in Figure~\ref{VS9W-BV}. The black arrow shows interstellar extinction vector for $A_v = 20$ mag, using Weingartner \& Draine (\cite{weingartner2001}) Milky Way model of $R_v = 3.1$. Lower panel: The same diagram as the upper panel, but only includes sources with known object-types.  Black lines shows the approximate boundaries of different types of objects: middle-left: carbon stars and OH/IR stars, middle right: PNe and post-AGB stars, bottom-left: M-type giants and supergiants, and S-type stars, bottom-right: YSOs, pre-main sequence stars, WR stars and Be stars (see text). }
\label{S9WL18W-JL18W}
\end{figure*}

\begin{figure*}
\centering
\includegraphics[angle=-90,scale=0.65]{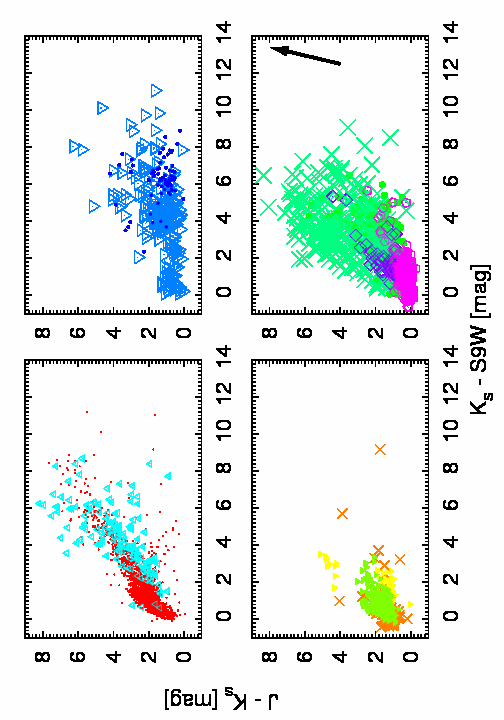}
\caption{The $(K_s-S9W)$ v.s. ($J-K_s$) color-color diagrams of sources with known object-types. Only sources with S/N $>$ 5 in both colors are included. Symbols are the same as in Figure~\ref{VS9W-BV}. The black arrow shows interstellar extinction vector for $A_v = 20$ mag, using Weingartner \& Draine (\cite{weingartner2001}) Milky Way model of $R_v = 3.1$.}
\label{KS9W-JK}
\end{figure*}

\begin{table*}
\caption{Number of sources in each region.}  
\label{table:nsource}
\begin{center}
\begin{tabular}{l r r r r r r r r r r r r}
\hline\hline
Region & \multicolumn{11}{c}{Known object-type} & Unidentified \\
     & Carbon & MIII  & MI      & S        & PMS   & YSOs  & OH/IR & PAGB  & PNe   & WR     & Be      &  \\
\hline
A   &    1441 &          1 &         0 &       37 &        0 &       11 &         6 &       15 &        1 &        9 &        2 & 9530 \\
B   &        27 &          4 &         0 &         5 &      19 &       88 &       90 &       44 &        9 &        0 &        5 & 14162 \\
C   &      548 &        75 &       50 &     344 &      58 &       45 &       27 &       20 &        0 &      12 &      11 & 62958 \\
D   &        36 &      812 &       12 &     117 &        1 &         0 &         0 &         3 &        0 &      10 &      92 & 23940 \\
E   &          4 &          0 &         3 &         1 &      72 &       35 &         2 &       19 &        1 &        0 &        6 & 1089 \\
F   &          6 &          0 &         0 &         0 &        6 &       19 &         2 &       61 &      65 &        2 &        3 & 1336 \\
\hline
\end{tabular}
\end{center}
\end{table*}

\subsubsection{ ($S9W-L18W$) v.s. ($J-L18W$)}
Figure~\ref{S9WL18W-JL18W} shows the ($S9W-L18W$) v.s. ($J-L18W$) color-color diagram using only infrared wavelength bands. We found more cross-identifications between AKARI-2MASS than between AKARI-Hipparcos. In this figure, not only nearby Hipparcos sources, but also more distant sources in the Milky Way, and some extragalactic point sources (e.g., red supergiants in the LMC) are present. However, the vast majority of the sources should be galactic, and extragalactic ones do not affect any of the discussions below.

\paragraph{\bf{Red giants}\\}
Carbon stars and oxygen-rich stars (M-type giants, supergiants, and OH/IR stars) are well separated in this diagram. 
This separation is mainly due to the different dust properties between carbon stars and oxygen-rich stars, although molecular bands are the main cause of the separation for relatively blue red giants. Carbon stars show a band attributed to SiC at 11.3\,$\mu$m, in addition to amorphous carbon dust, which has continuous emission without a characteristic feature in the infrared, and dominates dust mass and infrared excess in carbon stars (Groenewegen et al. \cite{groenewegen1995}). Carbon stars with higher mass-loss rate become red, due to amorphous carbon dust (Matsuura et al. \cite{matsuura2009}). The SiC feature falls in $S9W$-band (see Figure~\ref{filter}), and it is found in emission usually: only a few carbon stars are known to have the SiC feature in absorption (Speck, Barlow, \& Skinner \cite{speck1997}; Pitman, Speck, \& Hofmeister \cite{pitman2006}; Gruendl et al. \cite{gruendl2008}; Speck et al. \cite{speck2009}). Thus carbon stars tend to have a monotonous increase in both  ($S9W-L18W$) and  ($J-L18W$) color.

In contrast, heavily mass-losing oxygen-rich stars show the silicate bands at 9.8\,$\mu$m and 18\,$\mu$m. In Figure~\ref{S9WL18W-JL18W}, OH/IR stars contribute to the distinct sequence, amongst all oxygen-rich stars. Usually OH/IR stars have the silicate 9.8 $\mu$m feature in (deep) absorption, while the 18\,$\mu$m silicate band remains in emission (in exceptional cases in absorption; Sylvester et al. \cite{sylvester1999}). Therefore, OH/IR stars have slightly redder in ($S9W-L18W$) colors than those of carbon stars.

For relatively blue red giants, molecular features are the cause for locating carbon stars, M-type giants, supergiants, and S-type stars in different regions of the color-color diagram. Carbon stars have C$_2$ and CN absorption bands in $J$-band (Loidl et al. \cite{loidl2001}), suppressing $J$-band fluxes. In $S9W$-band, the wing of broad and strong C$_2$H$_2$ and HCN band 
is found (Matsuura et al. \cite{matsuura2006}). M-type giants and supergiants show CO and TiO absorption bands, and supergiants have additionally CN bands in $J$-band (Lan\c{c}on et al. \cite{lancon2007}), while M-type giants and supergiants have weak molecular features within $S9W$-band, such as CO$_2$ (Justtanont et al. \cite{justtanont1998}). In S-type stars, the major molecular features are CO, but weak features of carbon-bearing or oxygen-bearing molecules could be found, depending on the C/O ratio (Hony et al. \cite{hony2009}). All of these effects make characteristic colors of these groups.

Figure~\ref{KS9W-JK} is an another infrared color-color diagram for sources with known object-types similar to the Figure~\ref{S9WL18W-JL18W}. Only sources with S/N $>$ 5 in the colors in question are plotted. This figure demonstrates how the separations in the color-color diagram between the object-types would be if we place an emphasis on the near-infrared photometry. Now the separation between the carbon stars and the OH/IR stars is blurred, highlighting the importance of the AKARI data for the characterization of red giants. It can be emphasized that this diagram is useful to select mass-losing ($J-K > 4.0$) M-type supergiants from M-type giants.

\paragraph{\bf{Young stellar objects and pre-main-sequence stars}\\}
YSOs and PMS stars are well separated from AGB stars in the color-color diagram, although PNe and post-AGB stars are found to have similar colors. YSOs and PMS stars have lower luminosity in $L18W$-band than other types of stars with similar infrared excess. YSOs and PMS stars are marginally separable in Figure~\ref{S9WL18W-JL18W}, in the sense that the latter group has relatively bluer ($J-L18W$) color. Because we could not find criteria to distinguish YSOs from PMS stars in Wenger et al. (\cite{wenger2000}), here we assume that YSO are embedded Class I or II objects, and PMS stars represent more evolved Class III objects (Lada \& Wilking \cite{lada1984}). A marginal separation between the two groups is probably because the circumstellar disks become optically thin and the central stars become visible in PMS stars.

\paragraph{\bf{Unidentified objects}\\}
In the lower panel of Figure~\ref{S9WL18W-JL18W}, we defined six regions to make an approximate object classification on the color-color diagram. The representing object types in the regions are; A: Red carbon stars, B: OH/IR stars and some YSOs, C: M-type giants, supergiants and S-type stars, as well as bluer carbon stars in blue part, and PMS stars in red part, D: Be stars, M-type giants, and S-type stars, E: PMS stars, PNe, and PAGB stars, F: PNe and PAGB stars. Objects in regions A and B are most likely to be carbon stars and OH/IR stars, respectively, while PNe, post-AGB stars and YSOs tend to spread over the diagram, and it is difficult to identify these types of stars only with these infrared colors. Region D is mixed up with Be stars and red giants (M-type giants/supergiants and S-type stars). However, these two populations can be easily separated by seeing their optical color, such as ($B-V$) (see Figure~\ref{VS9W-BV}).

A comparison of the two panels of Figure~\ref{S9WL18W-JL18W} indicates that there are many infrared stars without object-type classifications. Some objects fall into regions A and B, and they are likely to be extremely red carbon stars and OH/IR stars, respectively. To explore these unidentified objects, we are now conducting follow-up observations, using the AKARI during post-helium mission, which take 2.5 -- 5 $\mu$m spectra with $\lambda / \Delta\lambda \sim 100$ for selected sources redder than $f_{\textrm{S9W}} / f_{\textrm{K$_s$}} > 2$ in $| b | > 30^\circ$ (P.I.: S. Oyabu) and also for selected sources in $| b | < 30^\circ$ (P.I.: D. Ishihara), where $f_{\textrm{S9W}}$ is $S9W$ flux in Jy,  $f_{\textrm{K$_s$}}$ is 2MASS's $K_s$-band flux in Jy and $b$ is the galactic latitude, respectively. In Table~\ref{table:nsource}, we summarize the number of sources with known object-type, and unidentified sources in each region.


\begin{figure*}
\centering
\includegraphics[angle=-90,scale=0.74]{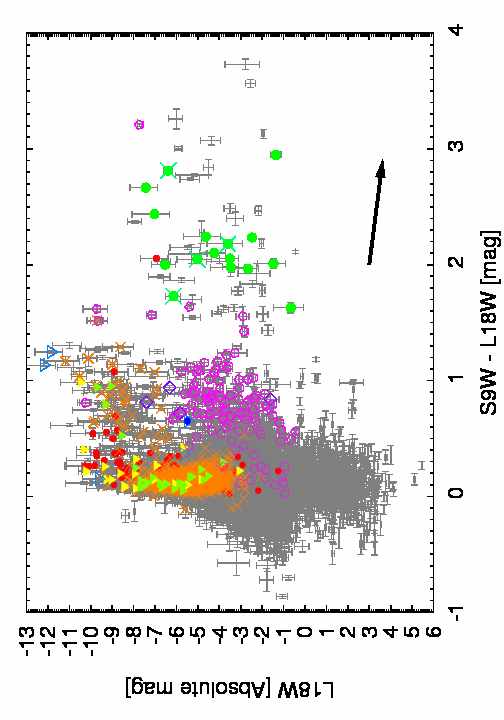}
\caption{The $(S9W-L18W)$ v.s. $M_{\textrm{L18W}}$ color-magnitude diagram. The error bars show $\pm$1 $\sigma$ errors. The vertical error bars include errors in parallax. Only sources with S/N $>$ 5 in color, and with $\sigma \omega/\omega <$ 0.4 are shown, where $\omega$ is the Hippparcos parallax. There are 13,252 sources that match the criteria. Symbols are the same as in Figure~\ref{VS9W-BV}. The black arrow shows interstellar extinction vector for $A_v = 20$ mag, using Weingartner \& Draine (\cite{weingartner2001}) Milky Way model of $R_v = 3.1$.}
\label{S9WL18W-L18W}
\end{figure*}

\begin{table*}
\caption{Young stellar objects or pre-main sequence stars plotted in the Figure~\ref{S9WL18W-L18W}.}  
\label{table:yso}
\begin{center}
\begin{tabular}{l r r r r l l l l l l}
\hline\hline
Common & R.A. & DEC & $S9W-L18W$ & $M_{\rm{L18W}}$ & Type & log $T_{\rm eff}$ &$L_{\star}$ & Mass & log (Age) & ref \\ 
name      & degree & (J2000) & (mag) & (abs. mag) & & (K) & ($L_{\odot}$) &($M_{\odot}$) & (yrs)\\ 
\hline
V773 Tau      & 63.55383    & 28.20341        & 1.97   & $-2.66$ & WT & 3.695 & 7.61 & 1.74 & 5.71  &b, c\\
T Tau             & 65.49764    & 19.53512        & 2.67   & $-7.43$ & TT & 3.708 & 14& 1.91 & 5.77 & b, c\\
DF Tau          & 66.76165    & 25.70620        & 1.63   & $-0.66$ & TT & 3.587 & 2.97 & 0.53 & 5.00 &b \\
GW Ori          & 82.28496    & 11.87019        & 2.44   & $-7.03$ & TT & & & & &b\\
HD  36112   & 82.61471    & 25.33252        & 2.00   & $-6.53$ & AB cand. & 3.91& 1.35 & 2.0 & 6.5 & a2 \\
CR Cha        & 164.77906  & $-77.02787$  &  2.18  & $-3.59$ & TT & & & & & b\\
TW Hya        & 165.46628  & $-34.70473$  & 2.95   & $-1.36$ & TT & & & & & b\\
DI Cha         & 166.83634  & $-77.63536$  &  2.05   & $-5.03$ & TT & & & & & b \\
CU Cha       & 167.01383  & $-77.65489$  & 2.81    & $-6.39$ & AB cand. & 4.00 & 1.61 & 2.5 & $>$ 6.3 & a2,b \\
HD 97300   & 167.45841  & $-76.61326$  &  1.73    & $-6.14$ &  AB         &   4.02     & 1.54  & 2.5   & $>$ 3 & a1 \\
HT Lup         & 236.30362  & $-34.29184$ & 2.10    & $-4.23$ & WT & 3.699 & 4.69 & 2 & 6--6.5 & e   \\
HD 141569  & 237.49062  & $-3.92121$  & 2.24    & $-2.47$ & AB cand. & 4.00 & 1.35 & 2.3 &$>$ 7.0 & a2 \\
RY Lup         & 239.86828  & $-40.36423$ & 1.98   &  $-3.47$ & TT & & & & & b \\
V1121 Oph  & 252.31377  & $-14.36906$ & 2.24   & $-4.63$ & TT & & & & & b \\
AK Sco         & 253.68687  & $-36.88849$ & 2.05   & $-3.51$ & AB cand./TT & 3.81 & 0.88 & & & a2, b\\
FK Ser         & 275.09479  & $-10.18712$  & 2.01   & $-1.47$ & TT & & & & & b \\
\hline
\end{tabular}
\end{center}
$L_{\star}$: luminosity of the central star,
Types: WT: Weak-lined T Tauri, TT: T Tauri, AB: Herbig Ae/Be, cand.:candidates \\ 
References:
a1: van den Ancker et al. (\cite{vandenancker1997}),
a2: van den Ancker et al. (\cite{vandenancker1998}),
b: Bertout, Robichon, \& Arenou (\cite{bertout1999}),
c: Beckwith et al. (\cite{beckwith1990}),
e: Gras-Velazquez \& Ray (\cite{gras2005}),
\end{table*}

\begin{figure}
\centering
\includegraphics[angle=90,scale=0.38]{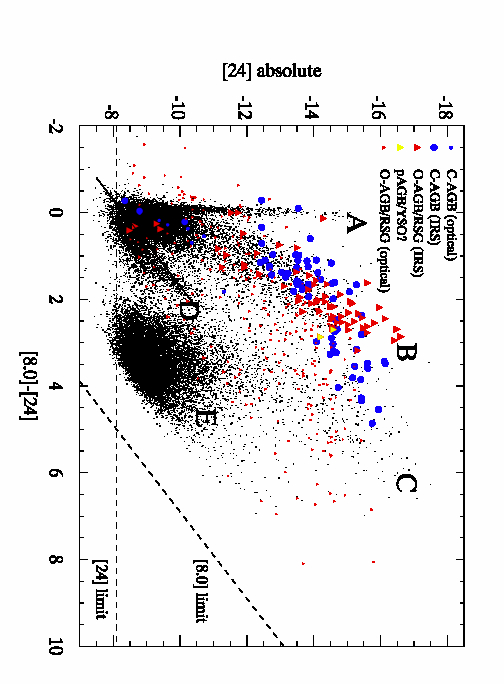}
\caption{The $([8.0] - [24])$ v.s. $M_{24}$ color-magnitude diagram of sources in the LMC. A distance modulus of 18.5 mag is adopted for the LMC sources. Symbols show oxygen-rich or carbon-rich AGB stars and red supergiants whose chemical types are spectroscopically confirmed. These previously known sources are taken from Matsuura et al. (\cite{matsuura2009}). Sequence 'A' is the foreground stars (not belong to the LMC). Sequence 'B' contains both oxgen-rich and carbon-rich AGB stars, and also contains heavily mass-losing stars at the tip ([24] $< -15$ mag). The solid line is defined by Srinivasan et al. (\cite{srinivasan2009}), and the sources below the line (sequence marked as 'D') are the fainter, redder O-rich population mentioned in Blum et al. (\cite{blum2006}), which correspond to red sequence in $(S9W-L18W)\sim1$ and $M_{\textrm{L18W}} \sim -9$. It appears both oxygen-rich and carbon-rich stars are found in sequence 'D' and 'B'. Dashed lines show detection limits of [8.0] and [24] quoted in Meixner et al. (\cite{meixner2006}).}
\label{i8m24colmag}
\end{figure}

\begin{figure}
\centering
\includegraphics[angle=-90,scale=0.36]{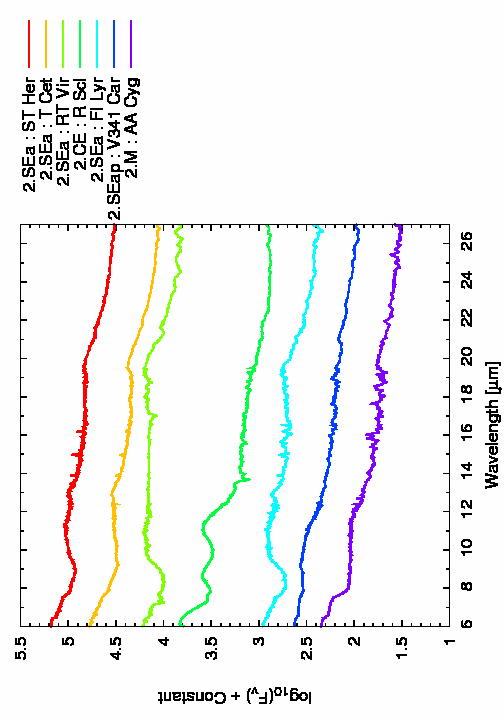}
\caption{The ISO/SWS spectra for all sources selected in section \ref{sectionfaint}. The scale of the vertical axis is arbitrary. Stars are sorted in increasing order of  $L18W$ luminosity from bottom to the top.}
\label{isosws}
\end{figure}

\begin{table}
\caption{These stars are classified as M-type supergiants in the SIMBAD, but  supergiants.}  
\label{supergiants}
\begin{center}
\begin{tabular}{l r r r }
\hline\hline
Name & $M_{\textrm{L18W}}$ & $(V-S9W)$ & $(B-V)$ \\ 
 & \multicolumn{3}{c}{[mag]} \\
\hline
KT Mus       & $-6.94$ & 6.50 & 1.77 \\
HD 306799 & $-4.85$ & 4.60 & 1.85 \\
RV Pup       & $-5.70$ & 6.79 & 1.41 \\
V408 Aur    & $-6.08$ & 5.45 & 1.79 \\
NSV 25773 & $-5.49$ & 4.40 & 1.69 \\
KN Cas       & $-3.08$ & 5.65 & 1.57 \\
\hline
\end{tabular}
\end{center}
\end{table}

\begin{table}
\caption{Galactic red giants that are possible counterparts to the fainter, redder giants in the LMC.}  
\label{faintgiant}
\begin{center}
\begin{tabular}{l l r r r}
\hline\hline
Name & Var. type  & Period [days] & $M_{\textrm{L18W}}$ & ISO/SWS \\ 
\hline
\multicolumn{5}{l}{\bf Carbon star} \\
R Scl                         & SRB & 370    & -8.96 & Y \\
TW Hor                     & SRB & 158    & -8.77 & N \\
Y Tau                        & SRB & 241.5 & -9.27 & N \\
V496 Car                   & SRB  & --      & -8.84 & N \\ 
HIP 56551                 & --       & --      & -6.94 & N \\
$^1$FI Lyr               & SRB & 146     & -8.91 & Y \\
U Cyg                       & Mira & 463.24 & -9.90 & N \\
\multicolumn{5}{l}{\bf S-type star} \\
T Cet                         & SRC & 158.9   & -9.32 & Y  \\ 
Y Lyn                         & SRC & 110      & -9.07  & N \\
ST Her                       & SRB & 148      & -9.71  & Y \\ 
AA Cyg                      & SRB & 212.7   & -8.57 & Y \\
\multicolumn{5}{l}{\bf M-type giants} \\
AC Cet                         & LB     & --          & -7.04 & N \\ 
V370 And                     & SRB  & 228       & -9.70 & N \\ 
Z Eri                              & SRB   & 80        & -7.69 & N \\ 
RR Eri                           & SRB    & 97       & -8.38 & N \\ 
VX Eri                           & SR:     & --         & -7.02 & N \\ 
SS Cep                         & SRB  & 90          & -8.72 & N \\ 
WX Hor                        & SRA  & --            & -8.51 & N \\ 
WW Pic                        & SRA  & --            & -6.67 & N \\ 
SW Col                         & LB:   & --            & -6.92 & N \\ 
ZZ Pic                           & LB     & --          & -6.97 & N \\ 
BQ Ori                          & SR     & --         & -6.98 & N \\ 
V341 Car                      & L      & --          & -8.61 & Y \\ 
AK Pyx                          & LB     & --          & -7.01 & N \\ 
RS Cam                        & SRB  & 88.6      & -6.72 & N \\ 
RT Cnc                          & SRB   & 60        & -8.06 & N \\ 
V489 Car                       & SRB  & --           & -8.34 & N \\ 
V496 Car                       & SRB  & --           & -8.83 & N \\ 
BB Ant                           & SRB  & 125        & -8.38 & N \\ 
T Crt                              & SRB   & --          & -6.65 & N \\ 
ST UMa                         & SRB  & 110       & -9.82 & N \\ 
AZ UMa                         & LB     & --          & -8.60 & N \\ 
Z UMa                           & SRB  & 195.5    & -8.52 & N \\ 
RW Vir                           & LB      & --         & -8.72 & N \\ 
RT Vir                            & SRB   & 155      & -8.98 & Y \\ 
V744 Cen                      & SRB  & --            & -8.38 & N \\ 
Y Cen                            & SRB:  & 180      & -9.21 & N \\ 
FY Lib                            & SRB   & 120      & -9.50 & N \\ 
R Ser                             & Mira   & 356.41 & -8.42 & N \\ 
V1293 Aql                      & SRB   & --         & -9.02 & N \\ 
V346 Tel                        & LB     & --            & -6.36 & N \\ 
V1070 Cyg                     & SRB   & 73.5     & -7.55 & N \\ 
W Cyg                            & SRB  & 131.1    & -9.08 & N \\ 
TW Peg                          & SRB   & 929.3   & -9.77 & N \\ 
BW Oct                           & LB:  & --           & -9.02 & N \\ 
DM Tuc                          & LB     & --          & -8.56 & N \\ 
Y Scl                              & SRB  &  --          & -7.68 & N \\ 
CC Ind                           & SRB:  & --          & -9.88 & N \\ 
S Phe                            & SRB   & 141       & -8.84 & N \\ 
\hline
\end{tabular}
$~^1$ this star can be a M-type star (see its ISO spectrum in Figure~\ref{isosws}).
\end{center}
\end{table}

\subsection{Color-magnitude diagram}
Nearby objects in the Galaxy have an advantage that we can make detailed study on them, while they may have a difficulty to correctly estimate the distance and thus to obtain the absolute magnitude. Objects in nearby galaxies, on the other hand, have similar distances and thus it is rather straightforward to make a color-magnitude diagram (CMD) for them. The Hipparcos data have changed the situation drastically and allow us to estimate the distance of nearby objects reliably. Combining the AKARI All-Sky Survey data with the Hipparcos data we are able to make a mid-infrared CMD for Galactic objects whose nature is well understood. Comparing the AKARI CMD with those of the LMC obtained by the Spitzer SAGE program (Meixner et al. \cite{meixner2006}) enables us to investigate populations in the LMC CMD, for which little information is available, such as "fainter, redder O-rich giants" (Blum et al. \cite{blum2006}, Srinivasan et al. \cite{srinivasan2009}).

\subsubsection{General overview}
Figure~\ref{S9WL18W-L18W} shows the $(S9W-L18W)$ v.s. $M_{\textrm{L18W}}$ infrared CMD, where $M_{\textrm{L18W}}$ is the absolute magnitude in $L18W$-band. Only sources with $\sigma \omega/\omega <$ 0.4 and S/N $>$ 5 in $(S9W-L18W)$ color are included in the figure, where $\omega$ and $\sigma \omega$ are the parallax and its error, respectively. There are 13,252 sources that matches these criteria. In our data set, the brightest stars in $M_{\textrm{L18W}}$ are post-AGB stars, followed by M-type supergiants and giants, and carbon stars and S-type stars. YSOs and PMS stars show large excess in the color $1.5 < (S9W-L18W) < 3$. Be stars tend to have lower $M_{\textrm{L18W}}$ luminosities with moderate $0.3 < (S9W-L18W) < 1.3$ excess, and WR stars also have similar magnitudes and colors as Be stars. M-type giants follow a sequence of $(S9W-L18W) \sim 0.1$ from $M_{\textrm{L18W}} = -3$ to $-8$, and these stars have little emission from the circumstellar envelopes. Once circumstellar envelopes are developed, M-type giants become redder in color. There may be two sequences in M-type giants. One sequence follows $M_{\textrm{L18W}} = -9$ mag up to  $(S9W-L18W) \sim 1.5$ and the other follows $M_{\textrm{L18W}} = -7$ mag up to  $(S9W-L18W) \sim1.2$. It is not clear whether these two sequences actually represent different populations or stars with different dust properties, or a continuous sequence with a large scatter due to complexity of dust and molecular features. Carbon stars tend to follow a similar trend as M-type giants with little excess. We have to interpret this CMD cautiously, as there is no parallax available for stars with heavily obscured central stars with circumstellar dust (i.e., heavily mass-losing infrared AGB stars).

\subsubsection{M-type supergiants}
There are six faint ($M_{\textrm{L18W}} >$ -7 mag) sources classified as M-type supergiants in Figure~\ref{S9WL18W-L18W}, namely KT Mus, HD 306799, RV Pup, V408 Aur, NSV 25773, and KN Cas. Their $M_{\textrm{L18W}}$, $(V-S9W)$ and $(B-V)$ values are tabulated in Table~\ref{supergiants}. Even if they are without circumstellar dust, they are still too faint to classify as supergiants. They are more likely to be M-type giants, judging from their L18W luminosities. Furthermore, all of them have $(V-S9W)$ and $(B-V)$ colors similar to the general colors of M-type giants (see section 3.1.1).

\subsubsection{Young stellar objects and pre-main-sequence stars}
YSOs and PMS candidates can be selected in Figure~\ref{S9WL18W-L18W}. In that figure, these types of stars are found at ($S9W-L18W$) $>$1, although some contaminations of other populations, such as post-AGB stars and PNe are expected. There are 16 YSOs or PMS stars in Figure~\ref{S9WL18W-L18W}. The number is mostly limited by the Hipparcos detection limit and parallax errors, and stars without apparent central stars in optical are not found in this diagram. Their common names, celestial coordinates, ($S9W-L18W$) colors, and $M_{\rm{L18W}}$ absolute magnitudes are listed in Table~\ref{table:yso}. 

A bibliographical survey shows that these stars are T Tauri and Herbig Ae/Be stars. It is clear that all of them show infrared excess, and the excess should originate from dust emission in their circumstellar disk (e.g., Whitney et al. \cite{whitney2003},  Adam et al. \cite{adam1987}). These stars are distributed in a relatively small range in ($S9W-L18W$) color (about 1 mag) while a wider range in $M_{\textrm{L18W}}$ luminosity (about 7 mag). Most of these stars are likely to have disks. The cause of the spread in luminosities  ($M_{\rm{L18W}}$) is not clear from the table, however, we suggest possibilities, such as the differences in the viewing angle of the disk (Adam et al. \cite{adam1987}), the inner radius of the disk,  and the disk mass. It appears that luminosities ($M_{\rm{L18W}}$) do not correlate with the stellar mass in our sample.

\subsubsection{Comparison with the Spitzer's color-magnitude diagram in the LMC}\label{sectionfaint}
AKARI's mid-infrared CMD helps understanding $([8.0] - [24])$ v.s. $M_{24}$ color-magnitude diagram of \textit{Spitzer} Space Telescope photometric data, such as those from Magellanic Clouds catalog (LMC: Meixner et al., \cite{meixner2006}, SMC: Gordon et al., \cite{gordon2010}).


We compare our $(S9W-L18W)$ v.s. $M_{\textrm{L18W}}$ diagram of galactic objects with the \textit{Spitzer} $([8.0] - [24])$ v.s. $M_{24}$ diagram of the LMC sources as shown in Figure~\ref{i8m24colmag}. After considering the offset values given in the Appendix, $(S9W-L18W) \sim 1$ should correspond to $([8.0]-[24]) \sim 1.7$. Therefore the galactic M-type giants, carbon stars and S-type stars with infrared excess ($0.4 < (S9W-L18W) < 1.5$ and $M_{\textrm{L18W}} < -6$ in absolute magnitude) probably correspond to the LMC fainter, redder sources located below the solid line (i.e., sources located on or below the sequence 'D') indicated in the Figure~\ref{i8m24colmag}\footnote{Note that the name of the sequences in Figure~\ref{i8m24colmag} are irrelevant to the region names defined in Figure~\ref{S9WL18W-JL18W}}, which is marked in Blum et al. (\cite{blum2006}) and Srinivasan et al. (\cite{srinivasan2009}). 

As the counterparts to the fainter, redder sources in the LMC, we look into the properties of the galactic less luminous (although they are among the brightest sample in our data set) red giants. We extract M-type giants, carbon stars, and S-type stars that satisfy ($S9W-L18W$) $>$ 0.4 and $M_{\textrm{L18W}} < -6$. There are 4 S-type stars, 7 Carbon stars, and 38 M-type giants that match the criteria. Then, we checked their pulsation properties (variability type and pulsation period), and also searched for their ISO/SWS spectra. The results are summarized in Table~\ref{faintgiant}. We found that all but one (HIP 56551) stars are known variable stars. Most of them show irregular or semi-regular type light variations. Judging from their relatively long pulsation periods, it is likely that they are on the asymptotic red giant branch (AGB), because faint variables with luminosities at around or below the tip of the first red giant branch (RGB) have shorter periods of about 30 days (e.g., Ita et al. \cite{ita2004}). Among the 49 samples listed in Table~\ref{faintgiant}, the ISO/SWS spectra (Sloan et al. \cite{sloan2003a}) are available for 7 stars. These spectra are shown in Figure~\ref{isosws} with their names and classification indices defined in Kraemer et al. (\cite{kraemer2002}). According to their classification, group 2 includes sources with SEDs dominated by the stellar photosphere but also influenced by dust emission. The SE and CE subgroups correspond to the oxgen-rich dust emission and carbon-rich dust emission, respectively. The M subgroup denotes miscellaneous. It is clear that all of the stars are surrounded by optically-thin circumstellar dust shells. Silicate dust features are seen in almost all M-type giants and S-type stars. Also, SiC feature at around 11.3 $\mu$m can be seen in carbon stars. Interestingly, so-called '13 $\mu$m feature' is seen in M-type and/or S-type giants. This feature is probably due to aluminum oxides (e.g., Posch et  al. \cite{posch1999}). Sloan et al. (\cite{sloan2003b}) suggested that this feature tends to be stronger in systems with lower infrared excesses and thus lower mass-loss rates (e.g., Onaka et al. \cite{onaka1989};  Kozasa \& Sogawa \cite{kozasa1997}). Interestingly, Sloan et al. (\cite{sloan1996}) reported that the 13 $\mu$m feature is preferentially detected in semi-regular or irregular variables. Based on these available data of galactic counterparts, we suggest that low mass-loss rate M-type giants, S-type stars, and carbon stars are mixed up to make sequence 'D' in the LMC (Figure~\ref{i8m24colmag}).

The counterparts to the sources on the brighter part ([24] $< -13$) of sequence 'B' are not present in AKARI $(S9W-L18W)$ v.s. $M_{\textrm{L18W}}$ color-magnitude diagram, due to the lack of reliable parallax measurements and/or such bright sources are saturated either in $S9W$ or $L18W$ measurements.

\section{Summary}
AKARI's mid-infrared All-Sky Survey increased the number of known mid-infrared sources drastically, mainly because of better spatial resolution than previous mid-infrared surveys. We combined the first release version ($\beta$-1) of the AKARI IRC All-Sky survey point source catalog with the existing all-sky survey catalogs, namely the Hipparcos and the 2MASS. Two-color diagrams are made with an aim to classify sources. We found that oxygen-rich giants and carbon stars are well separated by adding AKARI's new data. Also, we showed that Be stars and Wolf-Rayet stars with strong infrared excesses can be effectively selected by using optical and AKARI's combined colors. Combined with Hipparcos parallax measurements, we plot an infrared color-magnitude diagram. We uncovered the properties of redder, fainter red giants in the LMC, by comparing their galactic counterparts. This work will be greatly expanded in the forthcoming GAIA era, when we have good parallax measurements for tens of millions of stars. AKARI's new All-Sky Survey data reveal not only the mid-infrared characteristics of known objects, but also the existence of many yet-unidentified infrared sources. The color-color and color-magnitude diagrams we presented can be used to extrapolate the properties of the unidentified objects, invoking a follow-up campaign.

\begin{acknowledgements}
We thank the anonymous referee for the constructive comments that helped to improve this paper. This work is based on observations with AKARI, a JAXA project with the participation of ESA. This work is supported by the Grant-in-Aid for Encouragement of Young Scientists (B) No.~21740142 from the Ministry of Education, Culture, Sports, Science and Technology of Japan.This research has made use of the SIMBAD database, operated at CDS, Strasbourg, France. This publication makes use of data products from the Two Micron All Sky Survey, which is a joint project of the University of Massachusetts and the Infrared Processing and Analysis Center/California Institute of Technology, funded by the National Aeronautics and Space Administration and the National Science Foundation. This research has made use of the NASA/IPAC Extragalactic Database (NED) which is operated by the Jet Propulsion Laboratory, California Institute of Technology, under contract with the National Aeronautics and Space Administration.
\end{acknowledgements}

\begin{appendix}
\section{}
\begin{figure}
\centering
\includegraphics[angle=-90,scale=0.36]{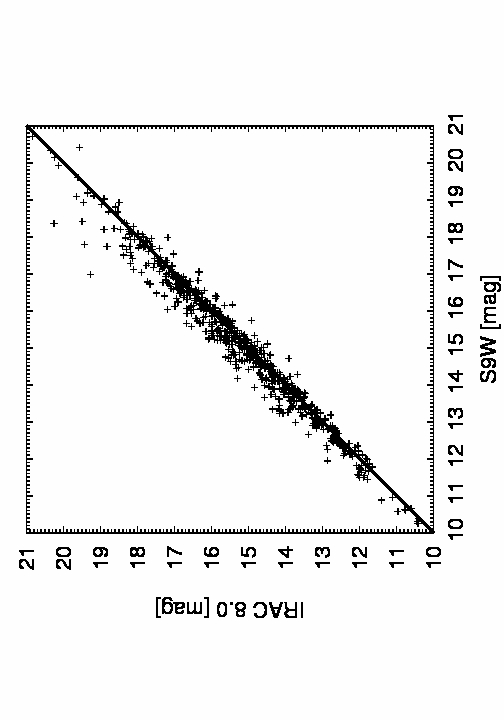}
\includegraphics[angle=-90,scale=0.36]{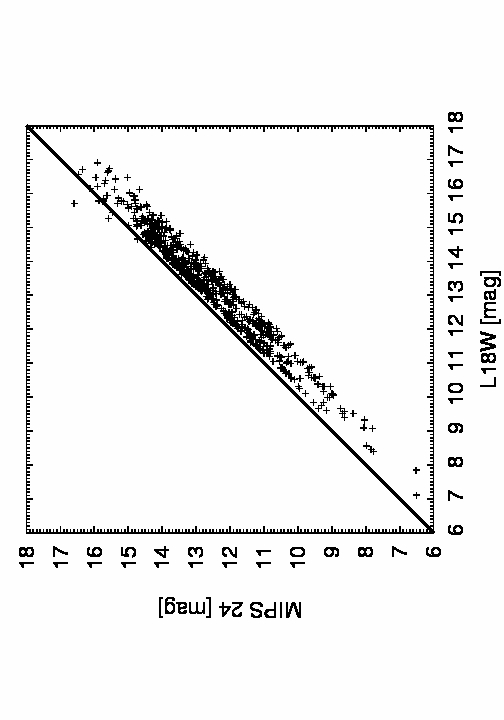}
\includegraphics[angle=-90,scale=0.36]{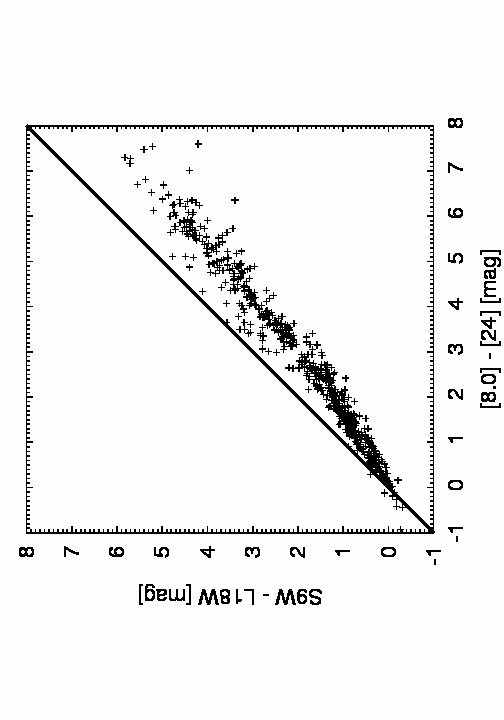}
\caption{The comparison of calculated magnitudes through synthetic photometry on ISO/SWS spectra for $S9W$ and IRAC 8.0 (upper panel), $L18W$ and MIPS 24.0 (middle panel), and colors (bottom panel).}
\label{comparison}
\end{figure}

We demonstrate conversion between the AKARI $(S9W-L18W)$ v.s. $M_{\textrm{L18W}}$ and the \textit{Spitzer} $([8.0] - [24])$ v.s. $M_{24}$ color-magnitude diagrams. Using the ISO/SWS flux-calibrated spectral library (Sloan et al. \cite{sloan2003a}), we calculate synthetic magnitudes of AKARI's $S9W$- and $L18W$-bands and Spitzer's IRAC $[8.0]$ (Fazio et al. \cite{fazio2004}) and MIPS $[24]$ (Rieke et al. \cite{rieke2004}). This is the same method as is written in Ita et al. (\cite{ita2008}). Figure~\ref{comparison} shows correlations between calculated $S9W$ v.s. $[8.0]$, $L18W$ v.s. $[24]$, and $(S9W-L18W)$ v.s. $([8.0]-[24])$. The relation between $S9W$ and $[8.0]$ is almost linear, with a median offset ($\equiv [8.0]-S9W$) value of 0.13 mag ($\sigma = 0.35$ mag). There is also a linear correlation between $L18W$ and $[24]$ with a notable offset ($\equiv [24]-L18W$) of $-0.62$ mag ($\sigma = 0.35$ mag). Actual transformations needs color corrections as found in the bottom panel of the figure, but general trend can be discussed even if we ignore the color correction terms.

\end{appendix}

\end{document}